\documentclass[twocolumn,prb,showpacs,amsmath,amssymb]{revtex4}

\usepackage{graphicx}
\usepackage{dcolumn}
\usepackage{bm}
\def\Mn12{Mn$_{12}ac$}
\def\cm{{\rm cm}$^{-1}$}

\begin{document}

\title{Asymmetric Lineshape due to Inhomogeneous Broadening \\ of the
Crystal-Field Transitions in \Mn12\  Single Crystals}

\author{S. Vongtragool$^{1}$,
 A. Mukhin$^{1,2}$, B. Gorshunov$^{1,2}$, and  M. Dressel$^{1}$} \affiliation{$^1$ 1.~Physikalisches
Institut, Universit{\"a}t Stuttgart,
Pfaffenwaldring 57, 70550 Stuttgart, Germany \\
$^2$ General Physics Institute, Russian Academy of Sciences, Moscow, Russia}

\date{\today}

\begin{abstract}
The lineshape of crystal-field transitions in single crystals of
\Mn12\ molecular magnets is determined by the magnetic history.
The absorption lines are symmetric and Gaussian for the
non-magnetized state obtained by zero-field cooling (zfc). In the
magnetized state which is reached when the sample is cooled in a
magnetic field (fc), however, they are asymmetric even in the
absence of an external magnetic field. These observations are
quantitatively explained by inhomogeneous symmetrical (Gaussian)
broadening of the crystal-field transitions combined with a
contribution of off-diagonal components of the magnetic
susceptibility to the effective magnetic permeability.
\end{abstract}


\maketitle

\section{Introduction}
The high-spin magnetic cluster of \Mn12\ (which stands for
[Mn$_{12}$O$_{12}$(CH$_{3}$COO)$_{16}$(H$_{2}$O)$_{4}$]$\cdot$2CH$_{3}$\-COOH $\cdot$ 4H$_{2}$O) is the
prime example of single-molecule based magnets which reveal a number of interesting phenomena like
mesoscopic quantum tunneling of the magnetization and quantum phase interference
\cite{Sessoli93,Gatteschi94,Friedman96,Barbara99,Caneschi99,Wernsdorfer99}. Over a wide range the
properties of \Mn12\ are determined by the splitting of the crystal field (CF) of the $S=10 $ ground
multiplet described by the spin Hamiltonian \cite{Barra97,Hill97,Hennion97,Mukhin98}
\begin{equation}
{\cal H} = DS_z^2 + D_4S_z^4 + B_4^4(S_+^4 + S_-^4)/2 - g\mu_B
{\bf S}\cdot{\bf H}  \label{eq1}
\end{equation}
where the first three terms represent the crystal field and the
last one the Zeeman energy in a  magnetic field ${\bf
H}$. The large single-ion anisotropy with an energy barrier of
$\sim 65$~K is produced mainly by the axial term  $D S_z^2$ with
$D<0$. The magnetization {\bf M} of the cluster is stabilized
along the fourth order $C_4$ symmetry axis. At low temperatures
resonant tunneling through the energy barrier is possible if the
energy levels on both sides coincide
\cite{Friedman96,Caneschi99,Barbara99}.

In the last years the structure of the energy levels of the ground
state multiplet of \Mn12\ and the parameters of the effective spin
Hamiltonian (\ref{eq1}) were studied by EPR measurements
\cite{Barra97,Hill97}, inelastic neutron scattering
\cite{Hennion97} and quasioptical magnetic spectroscopy
\cite{Mukhin98,Mukhin01,Mukhin02}. Currently much attention is
devoted to the shape and width of the absorption lines (CF
transitions) which may provide a deeper understanding of the
mechanisms of quantum tunneling in real crystals of molecular
magnets. The Gaussian lineshape found at zero field
\cite{Mukhin01,Mukhin02} directly indicates an inhomogeneous
character of the line broadening due to a distribution of
intra-crystalline interactions, in particular, magneto-dipolar and
crystal fields. Detailed EPR investigations
\cite{Parks01,Park02a,Hill02,Hill03} of various contributions to
the line broadening (distribution of the crystal fields,
$g$-factors, dipolar fields) revealed an important role of the CF
distribution coming from local strains ($D$-strain). In addition,
the asymmetry of the EPR lines was attributed \cite{Park02b} to
the distribution of the easy-axis orientation. In particular,
dislocations in real crystals were suggested to cause a
distribution of crystal fields which strongly influences the
mechanism of quantum tunneling \cite{Chudnovsky01}. We have
applied a novel type of magnetic spectroscopy to study the
influence of the inhomogeneous broadening on the lineshape of
magneto-dipolar transitions in \Mn12\ which allowed us to reveal
new features and to advance an alternative explanation.

\section{Experimental Results}
A few dozen of \Mn12\ single crystals of typically 1 to 2~mm in size were aligned to a mosaic of about
0.5~mm thickness and $7\times 7~{\rm mm}^2$ area such that the $C_4$ axes of the crystals lay in the plane
of the plate and were parallel to each other (to an accuracy better than 3$^{\circ}$). Using a
frequency-domain magnetic spectroscopy \cite{Kozlov98}, we studied the magnetic absorption by measuring
the optical transmission coefficient in the frequency range from 8 to 18~\cm. The experiments were
performed at temperatures $T = 2.33$~K in a magnetic field $\bf H$ up to 7~Tesla which was oriented
perpendicular to the magnetic field vector $\bf h$ of the radiation propagating with  the  wavevector
${\bf q}$ along the $y$-direction (Voigt geometry: ${\bf h} \parallel x$, ${\bf q}\parallel y$, ${\bf H}
\parallel z\parallel C_4$).

\begin{figure}
\includegraphics*[width=8cm]{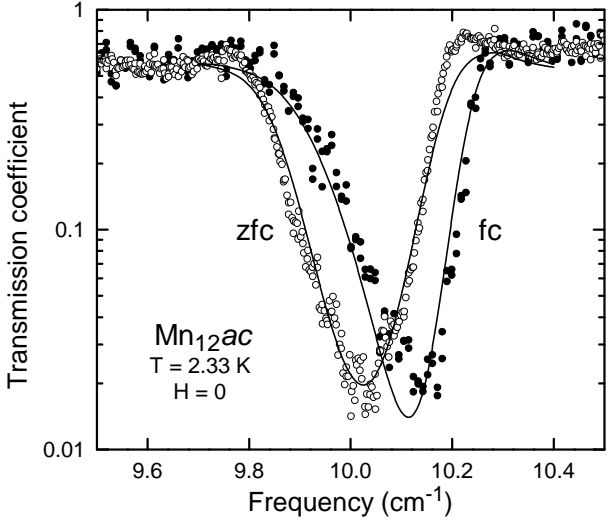}
\caption{\label{fig1}Magnetic absorption due to the $|\pm
10\rangle \rightarrow |\pm 9\rangle$ crystal-field transition
of \Mn12\ single crystals  as observed in the transmission spectra
at $T=2.33$~K without external magnetic field.
In one case (zfc) the sample is cooled down without applied
magnetic field, in the other case (fc) it is cooled in a magnetic
field of 1~T which was subsequently switched off. The solids lines
represent the calculations using a single Gaussian line and an
effective susceptibility as described in the text.}
\end{figure}
\begin{figure}
\includegraphics*[width=8cm]{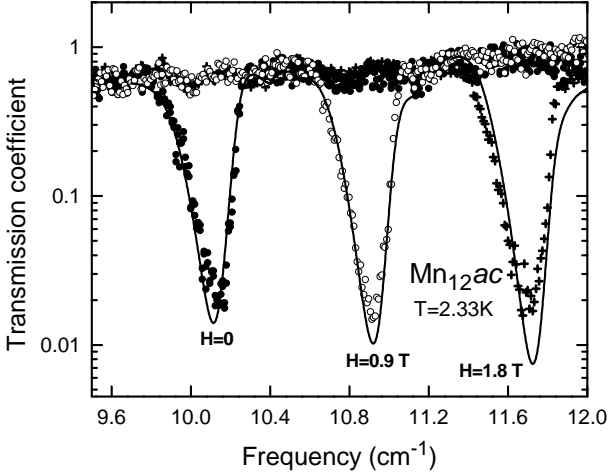}
\caption{\label{fig2}Absorption of the $|\pm 10\rangle \rightarrow |\pm 9\rangle$
crystal-field transition in \Mn12\ single crystals at $T=2.33$~K measured
by transmission at different magnetic fields. The lines represent
the calculations using a single Gaussian term and
an effective susceptibility.}
\end{figure}

The two transmission spectra plotted in Fig.~\ref{fig1} show the  $|\pm 10\rangle\rightarrow |\pm 9
\rangle$ transition measured at $T=2.33$~K with no external field present: $H=0$; note that this
corresponds to a real zero-field EPR experiment. In one case (non-magnetized, zfc state) the sample is
cooled down without applied magnetic field, in the second case (magnetized, fc state) the crystals are
cooled from 40~K in a field of 1~T (parallel to $C_4$) which was subsequently switched off.
In fc crystals
the resonance frequency is shifted up by 0.1~\cm\ and the lineshape is significantly asymmetric, with the
slope smoother on the left (low-frequency) side. Applying a magnetic field parallel to the $C_4$ axis of
the fc crystals increases the transition frequency as displayed in Fig.~\ref{fig2}; the lineshape remains
asymmetric similar to the $H=0$ case.

\section{Analysis}
For a quantitative description of the spectra we have to consider
the radiation propagating along $z$ axis through a transversely
magnetized medium ($h\|x, e\|y$ ), whose optical response is
determined by an effective magnetic permeability \cite{LaxButton}
\begin{equation}
\mu _{\rm eff}(\nu )
= \mu _{xx} (\nu ) - \mu _{xy} (\nu )\mu _{yx}
(\nu ) / \mu _{yy} (\nu ), \label{eq2}
\end{equation}
 and thus depends on both diagonal and off-diagonal components of the magnetic permeability
$\mu_{ij}(\nu)$ or susceptibility $\chi _{ik} (\nu )$ ($\mu _{ik}
(\nu ) = 1 + 4\pi \chi _{ik} (\nu )$) , where the latter are
sensitive to the magnetic state of the sample (fc or zfc). For low
temperatures only transitions from the ground state to the first
excited state $|\pm 10\rangle\rightarrow |\pm 9 \rangle$
contribute to the susceptibility
\begin{equation}
\chi_{xx,yy}(\nu) \equiv\chi_{\bot}(\nu ) =
\chi_{+10}R_{+10}(\nu)+ \chi_{-10}R_{-10}(\nu), \label{eq3}
\end{equation}
\begin{eqnarray}
\chi_{xy}(\nu)&=&-\chi_{yx}(\nu)\\
&=&i\nu[\chi_{+10}R_{+10}(\nu)/
\nu_{+10}-\chi_{-10}R_{-10}(\nu)/\nu_{-10}] , \nonumber
\label{eq4}
\end{eqnarray}
where $h\nu_{\pm 10} = E_{\pm 9} - E_{\pm 10} = h\nu_{10}^{0}\pm
g\mu_BH_z$ are the transition frequencies between the
corresponding states on one (+) or another (-) side of an
anisotropy barrier. Here $h\nu_{10}^{0}=-19[D+D_4(10^2+9^2)]$ is
the zero-field frequency and $\chi_{\pm 10}=2N\left(g\mu_B\langle
10|S_x|9\rangle\right)^2 \left(\rho_{\pm 10}-\rho_{\pm
9}\right)/{h\nu_{\pm 10}}$ are the contributions to the magnetic
susceptibility due to the corresponding transitions, $N$ is the
particle density,  and $\rho_{m}$ is the population of the energy
levels $E_m$ which are given by the  Boltzmann factor
$\rho_{m}=\exp\{-E_{m}/k_BT\}/ \sum\exp\{-E_{n}/k_BT\}$ in an
equilibrium state.
 The lineshape functions $R_{\pm 10}(\nu)\equiv R_{m}(\nu)$ may be
either Lorentzian R$_{m}(\nu )=\nu_{m}^{2}$/($\nu _{m}^{2}-\nu
^{2}$ + i$\nu \Delta \nu _{m})$, or Gaussian \cite{Mukhin01} with
the imaginary part
\begin{eqnarray}
R^{\prime \prime}_{m}(\nu )=(\pi /8)^{1 / 2}(\nu/\sigma_{m})
\left\{exp[-(\nu -\nu_{m})^{2}/2\sigma_{m}^{2}]  \right.
\nonumber \\
\left. + exp[-(\nu
+\nu_{m})^{2}/2\sigma_{m}^{2}]\right\},
\label{ImGauss}
\end{eqnarray}
and the real one determined via the
Kramers-Kronig relation
$R^{\prime}_{m}(\nu)=(2/\pi) \int \limits_0^\infty \nu _{1}
R^{\prime \prime}_{m}(\nu_{1})/(\nu_{1}^{2}-\nu ^{2}){\rm d}\nu_{1}$.

In a nonmagnetized state, when the population is the same on both sides of the barrier,
$\rho_{+10}\approx\rho_{-10}$, and the off-diagonal components of the susceptibility vanish
$\chi_{xy, yx}(\nu)= 0$, the effective permeability is given by
\begin{equation}
\mu _{\rm eff}^{\rm zfc}(\nu )
= \mu _{xx} (\nu ) = 1 + 4\pi \chi_{\bot}(\nu )
, \label{mu_zfc}
\end{equation}
where the diagonal susceptibility component $\chi_{\bot}(\nu )$ is determined by Eq.~(\ref{eq3}) for
$\chi_{+10}=\chi_{-10}\equiv \chi_{10}/2$ and $ R_{+10}=R_{-10}$.
In a magnetized state ($\rho_{+10}\approx$1,\ $\rho_{-10}\approx 0$)
\begin{eqnarray}
\mu _{\rm eff}^{\rm fc}(\nu )
&=&1 + \frac{4\pi \chi _{\bot}(\nu )[1+4\pi \chi
_{\bot}(\nu )(1-\nu^2/\nu_+^2)]}{1 + 4\pi \chi _{\bot}(\nu )}\nonumber \\
&\approx & 1 + \frac{4\pi \chi _{\bot}(\nu )}{1 + 4\pi \chi
_{\bot}(\nu )}, \label{mu_fc}
\end{eqnarray}
where $\chi _{\bot}(\nu) = \chi_{+10}R_{+10}(\nu)$. For $H=0$  the
functions $\chi _{\bot}(\nu)$ in the Eqs. (\ref{mu_zfc}) and
(\ref{mu_fc})  coincide and the difference
between zfc and fc permeability is mainly determined
by the denominator $1 + 4\pi \chi _{\bot}(\nu )$ in Eq.
(\ref{mu_fc}).

As shown in Fig.~\ref{fig1} by the solid lines, we can nicely
fit our experimental results for both the fc and zfc states by
using  the expressions for the transmission coefficient of a
plane-parallel layer \cite{DresselGruner} and a Gaussian lineshape
entering for $R_{\pm}(\nu)$ in $\chi _{\bot}(\nu )$.
 The main parameters, the
CF resonance frequency $\nu^0_{10} =10.002~{\rm cm}^{-1}$, the
Gaussian linewidth $\sigma_{10}^{\pm} =0.095~{\rm cm}^{-1}$
(corresponding full width at the half of maximum
$\Delta\nu_{10}^{\pm}=2\sqrt{2\ln2}\sigma_{10}^{\pm} \approx
0.224~{\rm cm}^{-1}$), and the contribution to the permeability
$\Delta\mu_{10}=4\pi\chi_{10}  = 0.076$, were found by fitting
only the zfc data \cite{remark1}. The calculated spectra reproduce
both, the upward shift of the resonance frequency and the
asymmetry of the lineshape in the fc state. The increase of the
resonant frequency from the value $\nu_{10}$ in zfc state to
 $\approx \widetilde{\nu}_{10}\sqrt{(1+\Delta\mu_{10})}$ in the fc
state is mainly due to the change of the effective permeability
since the off-diagonal term of the susceptibility becomes
significant. A  slight renormalization of the resonance frequency
$h\widetilde{\nu}_{10}=h\nu_{10} + g\mu_B \lambda_{\parallel} M_0$
in the fc state is caused by the weak internal dipolar magnetic
field; from our fit we obtain $\lambda_{\parallel}M_0 \approx 265
$ Oe (i.e.\ $g\mu_B\lambda_{\parallel}M_0 \approx 0.024~{\rm
cm}^{-1}$), where $M_0$ is the magnetization \cite{remark2}. The
above statement concerning the shift of the resonance frequency in
the fc state is true both for Lorentzian and Gaussian lineshape.
For the change of the shape, however, it is essential that the
lines are inhomogeneous broadened and have to  be described by
Gaussian lines.

\begin{figure}[b]
\includegraphics*[width=7cm]{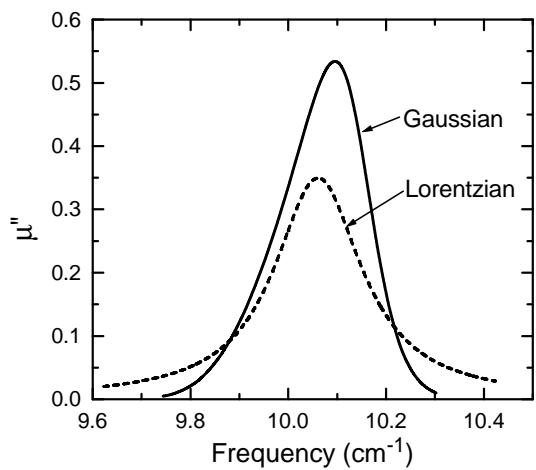}
\caption{\label{fig3} Magnetic absorption spectra $\mu^{\prime\prime}$
at zero magnetic field ($H=0$) for the field-cooled (fc) state of \Mn12.
The calculation compare Gaussian with Lorentzian lineshapes of the
$|\pm 10\rangle \rightarrow |\pm 9\rangle$
crystal-field transition.}
\end{figure}
In Fig.~\ref{fig3} the imaginary part of the permeability  spectrum
$\mu^{\prime\prime}_{\rm eff}(\nu)$ (which describes the magnetic
absorption) in the fc state is calculated assuming a Gaussian and a
Lorentzian lineshape $R_{m}(\nu)$.
It is clearly seen that the
line becomes asymmetric only in the case of the Gaussian lineshape
while for the Lorentzian case it remains symmetric.
This important result is connected to the fact that for a Lorentz\-ian lineshape
$R_{m}(\nu)$ in the Eq.~(\ref{mu_fc}) 
only the resonant frequency and the contribution (''strength'')
are renormalized,  while the overall frequency dependence of
$\mu_{\rm eff} (\nu )$ remains unchanged, i.e.\ Lorentz\-ian
\cite{remark4}. In the case of inhomogeneousely broadened
(Gaussian) lines, their shape is disturbed \cite{remark3}. Our
approach also explains the asymmetric lineshape for finite
magnetic fields, as demonstrated in Fig.~\ref{fig2}  where the
solid lines correspond to calculations using $g=1.93$ and the
aforementioned parameters.

\section{Discussion}
An asymmetrical EPR lineshape of \Mn12\ magneto-dipolar
transitions was reported in \cite{Hill02} and was explained
subsequently \cite{Park02b} by random distributions not only of
the $D$ parameter of the crystal field but also of the easy-axis
orientation. In this case, however, the asymmetry should vanish in
zero magnetic field which is in contradiction to our present
observations. Hence, the asymmetry of the lineshape can in
principle not be connected with the distribution of the easy axis
but rather is determined by a distribution of the zero-field
resonance frequencies. It should be noted that the experimental
measurement geometry (Voigt configuration) is of crucial
importance \cite{remark3}. Since regular EPR experiments are
performed at fixed frequencies as a function of applied magnetic
field --~in contrast to our experiments which are conducted at
fixed (finite or zero) magnetic field as a function of
frequency~-- the smaller slope is found on the right (high-field)
side relative to the resonance field as reported by Hill et al.
\cite{Hill02}. Assuming that the main contribution to the observed
linewidth is given by the dispersion of the crystal field $\Delta
D$ and taking into account that $\sigma_{10}^{\pm}/\nu_{10} =
\Delta D/(D+181D_4)\approx \Delta D/D$, we obtain $\Delta D
\approx 0.01D$ which is half the value estimated in \cite{Hill02}.

Recently it was shown \cite{Cornia02} that a disorder of the
acetic acids of crystallization induces different CF distortions
at the Mn$^{3+}$ ions positions; as a result six different isomers
can be found. The estimated CF parameters $D(n)$ for the most
populated isomers are \mbox{-0.769}~K, \mbox{-0.778}~K and
-0.788~K for $n=1,2,  3$, respectively  \cite{Cornia02}. The
relative differences $|D(2)- D(1)|/ D(2)= 0.0115$ and $|D(3)-
D(2)|/D(2)= 0.0128$ agree very well with the CF distribution obtained
from the linewidth. A simulation of the observed fc and zfc
spectra by three narrower Gaussians lines, corresponding to the
three isomers with the given concentrations, provides an excellent
description of the whole lineshape (not shown).  According to
Cornia et al.\ \cite{Cornia02} the CF distortions in the \Mn12\
isomers are accompanied by an appearance of a second-order
transverse CF term $E(S_x^2 - S_y^2)$ which is responsible for the
tunnel splitting and increase of the relaxation rate. The
correlation between $D$ and $E$ terms opens a possibility to
observe an inhomogeneous relaxation of the spectra inside the line
in a long-living non-equilibrium state created by magnetic field
inversion \cite{Dressel03}.

In summary, using high-frequency magnetic spectroscopy we
discovered a dependence of the frequency and lineshape of the
magneto-dipolar transitions in \Mn12\ on the magnetic history of the
sample. Zero-field cooled samples show a symmetrical Gaussian
lineshape of the $S=10$ ground multiplet transition $|\pm
10\rangle\rightarrow|\pm 9\rangle$. If the sample was previously
cooled in magnetic field, the line becomes asymmetric and shifts up in frequency by about
0.1~\cm\ even for $H=0$. These observations are explained by an inhomogeneous
(Gaussian) distribution of the resonant frequencies taking into
account the  effective susceptibility of the transversely
magnetized medium which includes off-diagonal components and the
internal magnetic field. We quantitatively describe the absorption
seen in the transmission spectra for both $H=0$ and $H\neq 0$ and
determined the main characteristics of interactions in the system.

\begin{acknowledgments}
 We thank the Prof.\ N. Karl and the Materials Lab for synthesizing the \Mn12\ single
crystals. We acknowledge helpful discussions with N.S. Dalal and J. Tejada.  This work was support by the
Deutsche Forschungsgemeinschaft (DFG) and by RFBR (No. 02-02-16597).
\end{acknowledgments}


\begin{thebibliography}{99}
\bibitem{Sessoli93} R. Sessoli, D. Gatteschi, A. Caneschi, and H.A. Novak,
Nature {\bf 356}, 141 (1993).
\bibitem{Gatteschi94} D. Gatteschi, A. Caneschi, L. Pardi, and R. Sessoli,
Science {\bf 265}, 1054 (1994).
\bibitem{Friedman96} J.R. Friedman, M.P. Sarachik, J. Tejada, and R. Ziolo,
Phys.\ Rev.\ Lett.\ {\bf 76}, 3830 (1996).
\bibitem{Barbara99} B. Barbara, L. Thomas, F. Lionti, I. Chioresu, and A. Sulpice,
J.\ Magn.\ Magn.\ Mat.\ {\bf 200}, 167 (1999).
\bibitem{Caneschi99}
 A. Caneschi, D. Gatteschi, C. Sangregorio, R. Sessoli, L. Sorace, A. Corina, M.N. Novak,
 C. Paulsen, and W. Wernsdorfer, J. Magn. Magn. Mat. {\bf 200}, 182 (1999).
\bibitem{Wernsdorfer99} W. Wernsdorfer and  R. Sessoli,
Science {\bf 284}, 133 (1999).
\bibitem{Barra97}
A.L. Barra, D. Gatteschi, and R. Sessoli, Phys. Rev. B {\bf 56},
8192 (1997).
\bibitem{Hill97}
S. Hill, J.A.A.J. Perenboom, N.S. Dalal, T. Hathaway, T. Stalcup,
and J.S. Brooks, Phys.\ Rev.\ Lett.\ {\bf 80}, 2453 (1997).
\bibitem{Hennion97}M. Hennion, L. Pardi, I. Mirebeau, E. Suard, R. Sessoli, and
A. Caneschi, Phys.\ Rev. \ B {\bf 56}, 8819 (1997); I. Mirebeau,
M. Hennion, H. Casalta, H. Andres, H.U. Gudel, A.V Irodova, and A.
Caneschi, Phys.\ Rev.\ Lett.\ {\bf 83}, 628 (1999).
\bibitem{Mukhin98}A.A. Mukhin, V.D. Travkin, A.K. Zvezdin, S.P.
Lebedev, A. Caneschi, and D. Gatteschi, Europhys.\ Lett.\ {\bf
44}, 778 (1998).
\bibitem{Mukhin01}
A.A. Mukhin, B. Gorshunov, M. Dressel, C. Sangregorio, and D.
Gatteschi, Phys.\ Rev.\ B {\bf 63}, 214411 (2001).
\bibitem{Mukhin02}
A.A. Mukhin, A.S. Prokhorov, B. Gorshunov, A.K. Zvezdin, V.D.
Travkin, and M. Dressel, Physics – Uspekhi {\bf 34}, 1306 (2002).
\bibitem{Hill03}S. Hill, R.S. Edwards, S.I. Jones, N.S. Dalal, and J.M. North, Phys.\ Rev.\ Lett.\ {\bf 90}, 217204 (2003).
\bibitem{Parks01}
B. Parks, J. Loomis, E. Rumberger, D.N. Hendrickson, and G.
Christou, Phys.\ Rev.\ B, {\bf 64}, 184426 (2001).
\bibitem{Park02a}
K. Park, M.A. Novotny, N.S. Dalal, S. Hill, and P.A. Rikvold,
Phys. Rev. B {\bf 65}, 014426 (2002).
\bibitem{Hill02}
S. Hill, S. Maccagnano, K. Park, R.M Achey, J.M. North, and N.S.
Dalal, Phys.\ Rev.\ B {\bf 65}, 224410 (2002).
\bibitem{Park02b}
K. Park, M.A. Novotny, N.S. Dalal, and P.A. Rikvold, J.\ Appl.\
Phys.\ {\bf 91}, 7167 (2002); K. Park, M.A. Novotny, and N.S.
Dalal, J.\ Chem.\ Phys.\ {\bf 117}, No.24 (2002).
\bibitem{Chudnovsky01}
E.M. Chudnovsky and D.A. Garanin, Phys.\ Rev.\ Lett.\ {\bf 87},
187203 (2001);\ D.A. Garanin and E.M. Chudnovsky, Phys.\ Rev.\ B
{\bf 65}, 094423 (2002); F. Torres, J.M. Hernandez, E. Molins, A.
Garica-Santiago, and J. Tejada,\ cond-mat/0110538.
\bibitem{Kozlov98} G.V. Kozlov and A.A. Volkov, {\em
Millimeter and Submillimeter Wave Spectroscopy of Solids}, ed. by G. Gr\"uner  (Springer, Berlin, 1998),
p.\ 51.
\bibitem{LaxButton}
See for example: B. Lax and  K.J. Button, {\em Microwave ferrites and ferrimagnets} (McGraw-Hill, New
York, 1962).
\bibitem{DresselGruner}
M. Dressel and G. Gr\"uner, {\em Electrodynamics of Solids} (Cambridge Univ. Press, Cambridge, 2002).
\bibitem{remark1}
The linewidth agrees well with the one obtained in polycrystalline \Mn12\
\cite{Mukhin98,Mukhin01,Mukhin02}. The found value of $\Delta\mu_{10}$ is slightly smaller than the
theoretical value 0.0109 obtained by $N=\rho N_A/M_{\rm Mn12}$ using the reported density $\rho=1.84~{\rm
g/cm}^{3}$ with $N_A$ the Avogadro number and $M_{\rm Mn12}$ the mass of the \Mn12\ molecule. The reason
may be the mosaic structure of the sample used in our experiments.
\bibitem{remark2}
Strictly speaking, also the transverse internal dipolar fields $\sim\lambda_{\bot}\Delta M_{\bot}$ may
become important when the magnetization deviates from the easy axis during oscillations; in general, this
effect could lead to a renormalization of  $\mu _{\rm eff} (\nu )$. However, our simulation show that this
effect is not relevant in our case.
\bibitem{remark4}
It can be easy checked by a direct substitution of $\chi _{\bot}(\nu) = \chi_{+10}R_{+10}(\nu)$ in
Eq.~(\protect\ref{mu_fc}).
\bibitem{remark3}
In Faraday geometry for a circular polarization the inhomogeneously broadened lines stays symmetric due to
another kind of effective permeability. These experiments are in progress.
\bibitem{Cornia02}
A. Cornia, R. Sessoli, L. Sorace, D. Gatteschi, A.L. Barra, and C.
Daiguebonne, Phys. Rev.\ Lett.\ {\bf 89}, 257201 (2002).
\bibitem{Dressel03} M. Dressel, B. Gorshunov, K. Rajagopal, S. Vongtragool, and A.A. Mukhin, Phys. Rev. B {\bf 67}, 060405(R) (2003).
\end{thebibliography}
\end{document}